%% file: conference_101719.tex
\documentclass[conference]{IEEEtran}
\IEEEoverridecommandlockouts
\usepackage{cite}
\usepackage{amsmath,amssymb,amsfonts}
\usepackage{algorithmic}
\usepackage{graphicx}
\usepackage{textcomp}
\usepackage{tabularx}
\usepackage{xcolor}
\usepackage{booktabs}
\usepackage{soul}

\def\BibTeX{{\rm B\kern-.05em{\sc i\kern-.025em b}\kern-.08em
    T\kern-.1667em\lower.7ex\hbox{E}\kern-.125emX}}
\usepackage[top=0.75in, bottom=1.05in, left=0.625in, right=0.625in]{geometry}
\begin{document}
\title{``Who Has the Time?'': Understanding Receptivity to  Health Chatbots among Underserved Women in India}

\author{\IEEEauthorblockN{\textsuperscript{} Manvi S}
\IEEEauthorblockA{\textit{Biomedical Informatics} \\
\textit{Emory University}\\
Atlanta, USA \\
mmanvi@emory.edu}
\and
\IEEEauthorblockN{\textsuperscript{} Roshini Deva}
\IEEEauthorblockA{\textit{Biomedical Informatics} \\
\textit{Emory university}\\
Atlanta, USA \\
droshin@emory.edu}
\and
\IEEEauthorblockN{\textsuperscript{} Neha Madhiwalla}
\IEEEauthorblockA{\textit{Research Director} \\
\textit{ARMMAN Organization}\\
Mumbai, India \\
neha@armman.org}
\and
\IEEEauthorblockN{}
\IEEEauthorblockA{} 
\and
\IEEEauthorblockN{\textsuperscript{} Azra Ismail}
\IEEEauthorblockA{\textit{Biomedical Informatics and }\\
\textit{Global Health} \\
\textit{Emory University}\\
Atlanta, USA \\
azra.ismail@emory.edu}}

\maketitle

\begin{abstract}
Access to health information and services among women continues to be a major challenge in many communities globally. In recent years, there has been a growing interest in the potential of chatbots to address this information and access gap. We conducted interviews and focus group discussions with underserved women in urban India to understand their receptivity towards the use of chatbots for maternal and child health, as well as barriers to their adoption. Our findings uncover gaps in digital access and literacies, and perceived conflict with various responsibilities that women are burdened with, which shape their interactions with digital technology. Our paper offers insights into the design of chatbots for community health that can meet the lived realities of women in underserved settings.


\end{abstract}


\input{introduction.tex}

\input{related_work}
\input{methods}

\input{findings}

\input{future_work}

\section{Conclusion}

Our study findings highlight the significant challenges and opportunities in using technical tools like chatbots to improve healthcare knowledge and accessibility for women and children. Despite the widespread smartphone and WhatsApp usage, participants face challenges in using technology due to time constraints, household responsibilities, and limited digital literacy.

Gaps in child health care knowledge were also very prominent among our participants. Participants expressed a clear desire for accessible, reliable information through chatbots, especially concerning their children's health and development. Our findings also revealed a preference for audio and visual content over text-based information, as many women faced challenges in reading and typing in Hindi or English. This suggests that future digital health interventions, including chatbots, should incorporate features such as voice messaging and visual aids to improve usability and engagement.

Overall, these findings emphasize the importance of designing culturally sensitive, user-friendly technological solutions that take into account the sociocultural constraints, digital literacy levels, and daily realities of women in these communities. By addressing these factors, chatbots and other digital tools have the potential to significantly improve healthcare access and outcomes for mothers and children.

\section*{Acknowledgment}

We want to extend heartfelt thanks to all participants who generously contributed their time and experience with us.  
We are also grateful to the NGOs who helped us recruit study participants and shared their knowledge and support to guide the practical application of digital technology for MCH. 
Thank you all for your invaluable contribution.

\vspace{12pt}

\end{document}

%% file: introduction.tex
\section{Introduction}

Maternal and Child Health (MCH) remains a critical concern in India, as it contributes to one-fifth of the global burden of maternal deaths \cite{b1}. Despite multiple interventions, significant gaps persist in delivering essential healthcare services and information to pregnant women and mothers regarding child care, nutrition, and immunization \cite{b2}. 
Recent studies suggest that digital technology, particularly chatbots, can help overcome these challenges by providing accessible and timely healthcare information \cite{b6,b7,22}. However, the adoption of digital technology for MCH remains uneven due to socioeconomic, cultural, and infrastructural barriers.

Our study focuses on understanding opportunities and challenges around the use of chatbots for MCH. 
We conducted 23 semi-structured interviews and two focus group discussions to understand how women in underserved communities in Mumbai (India) perceive chatbots and barriers to adoption. We examine the key concerns expressed by pregnant women and mothers, such as the complexities of healthcare access, digital literacies, cultural norms, and communication preferences. 
Based on our findings, we offer takeaways for the design of emerging health chatbots in global health that target women. In particular, we emphasize the need to consider intermittent digital access, support multiple modes of interaction, and consider women's everyday responsibilities and their limited agency in their households.

%% file: related_work.tex
\section{Related Work} 
Digital health technologies, including chatbots, have emerged as powerful tools with the potential to improve maternal health literacy, access to care, and overall health outcomes \cite{b6,b7,b17}, especially in developing countries \cite{b18}. Researchers on digital health solutions have leveraged widely accessible technologies \cite{b24}, including WhatsApp, for MCH information delivery \cite{b6}. For instance, Yadav et al. studied \cite{22} user interactions to deliver information to breastfeeding mothers of low-resource communities through a prototyped WhatsApp-based chatbot in India.
These innovations aim to improve access to health information and services, sometimes even offering integrated care, thus aiming to reduce health inequalities in underserved populations \cite{b9}. 
Despite many digital health efforts, however, there have been few sustained efforts and true successes. Evidence of the effectiveness of digital health for maternal and child health has been scarce \cite{b25}.

While smartphone access has increased significantly in underserved communities \cite{b4}, this has not translated into widespread adoption of digital health solutions \cite{b19}.
Many digital health initiatives have not sufficiently addressed access and engagement strategies, which are vital for their success to the populations most in need \cite{b11}. 
Studies on digital health also show that despite access to mobile phones, barriers such as technology-specific barriers \cite{20}, socio-cultural factors \cite{21}, digital literacy, and trust in technology \cite{b19} limit the effective use of healthcare technologies. 
Reaching the most vulnerable populations also entails careful consideration of contextual factors such as family dynamics, traditional health practices, and beliefs of the community \cite{22}. 
Even assuming that a digital health intervention has been designed well, its adoption at the last mile may depend on an individual's income level, education, and access to digital devices and the internet \cite{b10}.  
These disparities highlight the need for a more inclusive approach to technology development and deployment, ensuring that all socioeconomic groups can benefit equally. 
Our paper extends this body of work to consider implications for the design of chatbots for maternal and child health in underserved communities in India, even as we reflect on why these challenges still persist after almost two decades of research on digital health in the Global South. 
 


%% file: methods.tex
\section{METHODS}
Our study initially aimed to inform the design of a Whatsapp-based chatbot to provide maternal and child health (MCH) information to women in underserved communities. We chose WhatsApp due to its widespread reach in India, as well as its range of media offerings that feature voice, text, and video interactions. We are not unique in this experiment; this avenue is currently being tested by many health organizations in the hope of improving user engagement and accessibility. 
We conducted 23 semi-structured interviews and two focus groups with 15 women from underserved communities in Mumbai. 
We obtained approval to conduct the research from the Institutional Review Board at Emory University in the United States.

\subsection{Participant Recruitment}
Participants for this study were recruited through two different NGOs. For interviews, we recruited participants from NGO1 (anonymized), a non-profit based in Mumbai, India, that has been focused on maternal and child health since 2008. The participants were enrolled in the organization's child malnutrition program where they received calls from counsellors and automated voice messages regarding child nutrition and growth. The community workers from the organization conducted recruitment through phone calls and assisted in coordinating the interviews, which were held at the participants' homes. Our participant group reflects diversity in age, family status, education, number of children, and, annual income. Our study participants' demographics are shown in Table \ref{tab:demo}.

\begin{table}[htp]
\caption{Participants Demographics}
\begin{center}
\begin{tabular}{|l|c|c|}
\hline

\textbf{Attribute} & \textbf{N}& \textbf{\%} \\
\hline

    \textbf{Age} & & \\
    Age 18-29 & 17 & 73.9 \\
    Age 30-39 & 5 & 21.7 \\
    Age 40-49 & 1 & 4.3 \\   
\hline
    \textbf{Education} & & \\
    5th grade & 3 & 13.0 \\
    8th grade & 1 & 4.3 \\
    9th grade & 1 & 4.3 \\
    10th grade & 3 & 13.0 \\
    11th grade & 1 & 4.3 \\
    12th grade & 7 & 30.0 \\
    Diploma & 2 & 8.7 \\
    Graduation & 3 & 13.0 \\
    Post Graduation & 1 & 4.3 \\
    No education & 1 & 4.3 \\
  \hline
    \textbf{Housing situation} & & \\
    Own & 7 & 30.4 \\
    Rental, Nuclear family & 10 & 43.5 \\
    Rental, Joint family & 2 & 8.7 \\
    Ancestral, Joint family & 3 & 13.0 \\
    Ancestral, Nuclear family & 1 & 4.3 \\
  \hline
    \textbf{Annual household Income} & & \\
    Rs. 10,000-1,25,000 & 5 & 21.7 \\
    Rs. 1,26,000-10,00,000 & 2 & 8.7 \\
    Prefer not to say & 16 & 69.6 \\
\hline
    \textbf{Own Smartphone} & & \\
    Yes & 21 & 91.3 \\
    No & 2 & 8.7 \\
  \hline
    \textbf{WhatsApp Access} & & \\
    Yes & 23 & 100.0 \\
  \hline
\end{tabular}
\label{tab:demo}
\end{center}

\end{table}


In addition to these individual interviews, we conducted two focus group discussions (FGDs) with a total of fifteen participants. These participants were recruited from a second non-profit organization (NGO2) that provides support to mothers of children with disabilities. Both the focus groups were conducted in the communal space provided by NGO2 at their center.

\subsection{Data Collection and Analysis}
\begin{figure}[htp]
\centerline{\includegraphics[width=0.2\textwidth, height=0.3\textheight]{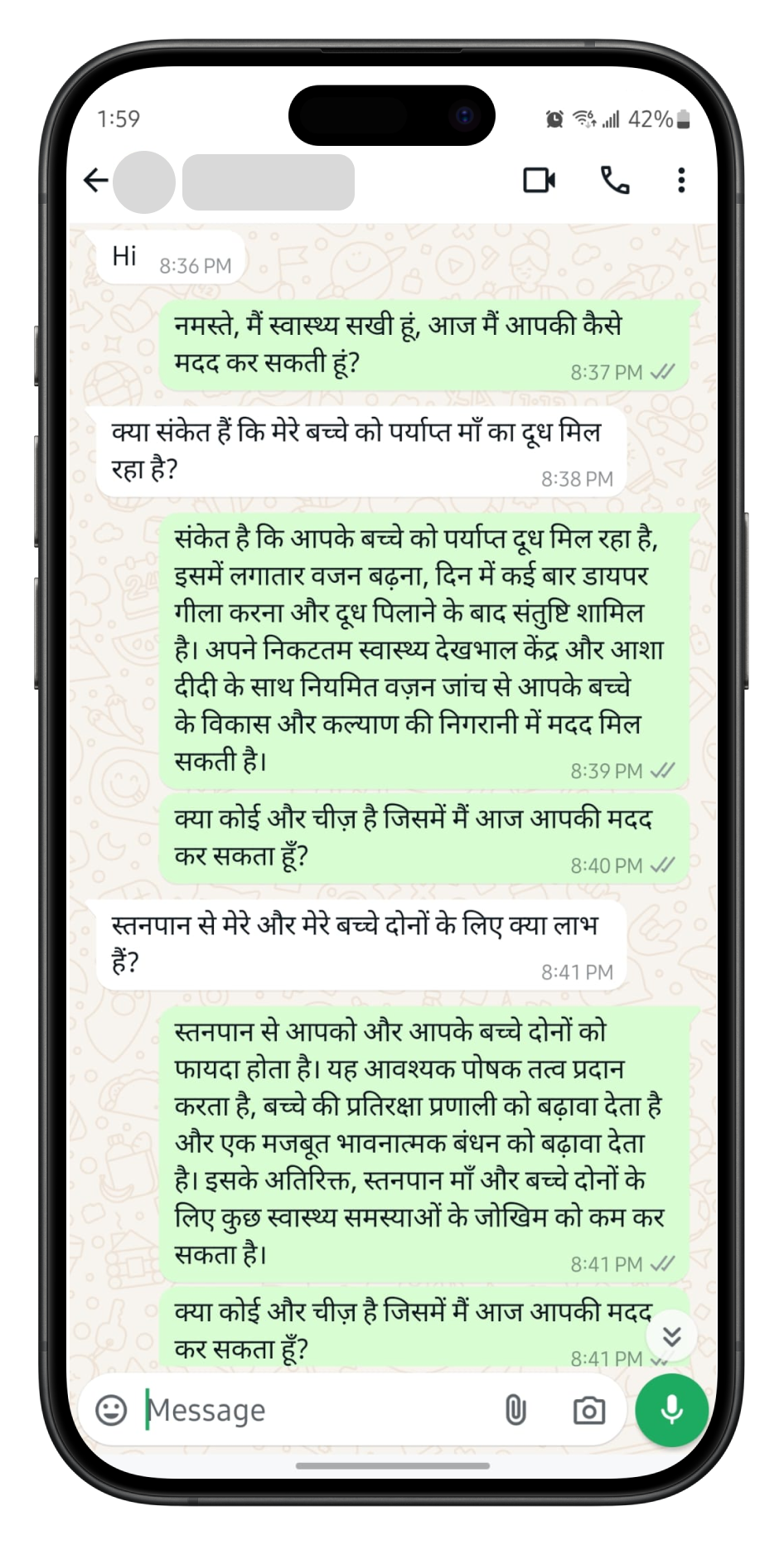}}
\caption{Sample Whatsapp conversation, Original Hindi text.}
\label{fig:whats1}
\end{figure}

We conducted semi-structured interviews with participants at their homes. With their consent, the conversations were recorded for further analysis. In addition to collecting demographic information, participants were shown a sample WhatsApp conversation in Hindi that included some common questions on breastfeeding and child nutrition (refer to Figure \ref{fig:whats1} and Figure \ref{fig:whats2}). Through the think-aloud protocol, a method used for usability testing \cite{b23}, we engaged participants with questions to understand their interest in technology, preferred modes of interaction, and perceptions of its usefulness.
Following the interviews, focus group discussions were conducted. These discussions were also recorded with the consent of participants. During the focus groups, an animated video clip on malnutrition, created by NGO1, was shown to the participants to prompt further discussion and understand their preferences for information. \\
Given that our participants spoke Hindi, we collected the data in Hindi. The data collected was then transcribed and translated into English for analysis. We then analyzed the data using a line-by-line coding approach to identify specific codes that represented participants’ responses, covering topics such as smartphone usage, familiarity with chatbots, preferred topics for chatbots, communication preferences, and family support in childcare. In the initial phase, coding closely followed the text of the transcripts.
Subsequently, we conducted a thematic analysis, grouping the codes into broader categories based on emerging patterns and themes. Through iterative refinement, we identified key thematic categories, including limited access to technology, information gaps in healthcare, the impact of socio-cultural norms, digital literacy, and trust in external programs and resources. To ensure the comprehensiveness of the data, thematic saturation was reached when no new themes emerged.

\subsection{Limitations}
To simulate chatbot interactions, our team initially planned to use the Wizard-of-Oz method, where participants would interact with an interface, and the responses are provided by a human operator. This approach has been successfully employed in prior studies in urban India to create the illusion of an AI interface \cite{22}. However, we faced practical constraints in executing this in an underserved community with intermittent internet, preventing real-time testing. 
Moreover, as the interviews progressed, it became evident that the Wizard-of-Oz method may not have been optimal given the participants' time constraints and the challenges they faced with text-based interactions.

To address this, we adapted our approach by developing a sample WhatsApp conversation in Hindi, which was presented to participants to elicit their preferences for technology use. While most participants were able to engage with the sample and offer valuable feedback, a few encountered difficulties reading the text in Hindi.

%% file: findings.tex
\section{Findings}
We now present the learnings from our analysis of interviews and focus group discussions (FGD).  Our findings identify the challenges limiting the MCH technology adoption among marginalized women. In several places, we have included quotes from participants in translated English text for the benefit of our readers. All names mentioned in the paper are pseudonyms.

\subsection{Limited access to technology}
We first summarize the patterns observed in barriers to phone usage among the participants. We then reflect on the phone sharing and access limitations that could present.

\subsubsection{Barrier to phone usage} 
Our interactions with mothers of children under 3 years of age highlighted significant time constraints. For instance, one of our participants shared her experience in the FGD1 discussion, ``\textit{I don’t get much time to use my phone due to household chores. I am busy most of the time, just 10 –15 minutes a day, that too when the child is sleeping''}. Another added, \textit{ ``I reply to a WhatsApp message if I get any for 5 -10min, other than that I don’t have the time to use my phone''}. These challenges significantly impact their ability to engage with smartphone applications, often limiting their usage to brief moments only when their children are napping or otherwise occupied. Even during the interviews, we observed that participants struggled to spend time due to their children needing attention. On the contrary, focus group participants were fully engaged in discussions without interruptions, likely because these conversations took place in a communal space, which may have created a more supporting setting. 
Given the limited time users can dedicate, the ability of the chatbots to provide asynchronous interactions can be useful. Our findings further indicated that the responses should also be precise and clear to facilitate prompt conversations. 

\begin{figure}[htp]
\centerline{\includegraphics[width=0.2\textwidth, height=0.3\textheight]{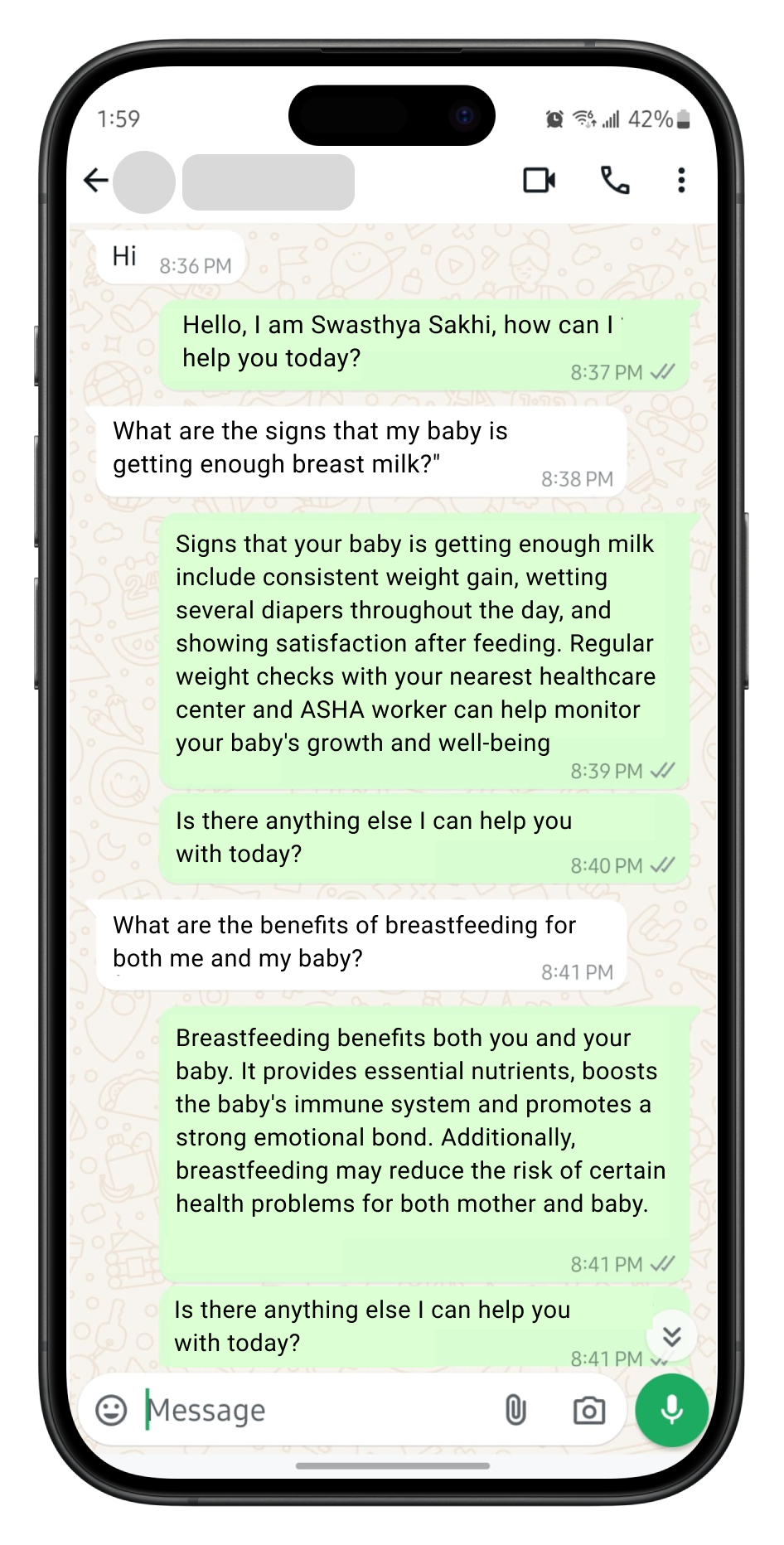}}
\caption{Sample Whatsapp conversation, English translated.}
\label{fig:whats2}
\end{figure}




\subsubsection{Phone sharing and access limitation}
We observed that participants faced challenges using technology, especially mothers like Poonam (28yr, a mother of 3 children), who relied on their husbands' phones and could only access phones during the evening. Chatbots may be more beneficial for users like Poonam if we include the men in the interventions. Taking privacy into account, chatbots can be designed with features like personalized to-do lists and the ability to set reminders at times that best suit the user, making it easier for them to use shared devices.

\subsection{Information gaps in childcare and health} 
We now share the observations on challenges in accessing information on childcare and health. We reflect on the implications of these gaps, especially due to the lack of accessible health information and the reliance on traditional sources, which may limit the participant's ability to make informed decisions.

\subsubsection{Lack of accessible health information} 
Our observations highlighted a gap in knowledge regarding child nutrition among mothers, with many indicating a need for guidance on appropriate dietary practices, particularly in special circumstances. One of our participants from FGD2 stated that: \textit{``My child goes to this daycare program in an NGO for disabled kids to learn new things, but it would be nice to know what should I feed him, what is good for his health?''}, which emphasizes the need for addressing nutritional gaps by providing accessible and tailored dietary information, empowering mothers, and allowing them to make informed choices that can improve their child's health outcome. 
Our findings indicate the need for chatbots to avoid generic nutritional information, as mothers might require personalized guidance on their children's needs. Chatbots should further consider collecting user specifics like age, dietary restrictions, and health concerns to provide tailored meal suggestions. Culturally specific information based on local food options makes the recommendations more practical, while partnerships with programs like NGO1 can connect mothers to helpful resources.

\subsubsection{Reliance on traditional sources}
Our analysis showed that while many mothers had access to smartphones and technology, they often relied on family or elders for healthcare advice rather than seeking information online. For instance, one of the mothers from FGD2 (mother of 2 young children) shared,\textit{`` I used to call my mother if I had any questions during my pregnancy or breastfeeding. It's good to have prior knowledge about child growth. When I had my 1st child, I was not aware of the child’s growth milestones. My first child did not speak until he was 3 years old, I was not aware that it was not normal, everybody around me said don’t worry he will speak. I did not know that a child should start saying words after 6 months. I had no idea that I should go to a doctor when a child does not start speaking by the age of 1 year''}. This implies the disparity between having access to technology and utilizing it for reliable health information. Chatbots can help address this gap and reduce dependence on potentially inaccurate information provided by the family. 



\subsection{Impact of social-cultural norms}
We now summarize observations on the challenges limiting technology use influenced by social and cultural norms such as gender roles and family dynamics. 

\subsubsection{Influence of family dynamics} 

Many of our participants noted that their decisions are heavily influenced by in-laws or husbands. In her interview, Seema (26 yrs, a mother of a 3-year-old boy) shared that: \textit{``We seek advice from the elders at home first and then go to the doctor if needed. The family members do have a say in decisions regarding health, when living with in-laws we have to get permission and tell them where and when we are going, and why are we going out''}. 
In analyzing the impact of family dynamics on health and technology decisions, it became evident that cultural beliefs do play a significant role in shaping women's choices. 




\subsubsection{Gender role limiting tech usage} 
The majority of participants in our study identified as stay-at-home mothers, whose primary responsibilities were taking care of the children and household management. This adherence to conventional gender roles not only restricts their time and chances to interact with technology but also limits their overall digital proficiency. Our data further reflects that the gendered division restricts women's access to technology and strengthens social hierarchies within the household, resulting in greater dependence on husbands or family members for decisions about technology.


{\subsection{Digital Literacy}}
Digital literacy also influenced participants' preferences for content, and frequently intersected with challenges related to general literacy. 
For instance, Nisha and Seema struggled with reading and writing Hindi or English. 
Their experience 
highlighted significant barriers to accessing health information through chatbots that primarily involve text in these languages. Nisha (24 years mother of 2 young children), who lived with her in-laws in a family of 10, mentioned that she has her smartphone with WhatsApp but has difficulty reading in Hindi or English, so she relied on her niece and husband. Their reliance on family for help not only delayed their access to information but also undermined their confidence in using technology on their own.

As shown in Figure 3, the majority of participants
expressed a strong preference for audio-based interactions.
This preference highlights the potential of audio-enabled functionalities to address barriers for users facing reading and
writing challenges. Furthermore, such features can enhance
usability for users managing other tasks while interacting with
the chatbot.

\begin{figure}[htp]
\centerline{\includegraphics[width=0.4\textwidth]{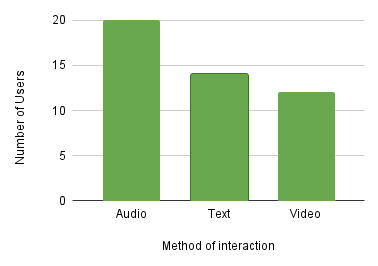}}
\caption{Interaction Preferences of the participants.}
\label{fig:fig}
\end{figure}

\subsection{Trust in external programs and resources}

Poonam (28 years old, a mother 3 children) noted that: \textit{``I do not like reading much, I like the weekly calls from the NGO for my son, as I do not get much time to use my phone, I am busy taking care of kids and housework''}. 
This indicated a strong opportunity to integrate chatbots into health organizations that support maternal and child health. As these programs were specifically designed for malnourished children, they were highly valued by the participants because of their easy accessibility and tailored messages. Other than the reliance on these programs and family members for health care advice, all participants mentioned clear reliance on doctors in case of serious medical issues. \\
All participants reported regular doctor visits during pregnancy, demonstrating a strong reliance on medical professionals for prenatal care. In response to the questions about iron and calcium supplements during pregnancy and family planning measures, 50\% of the participants responded ``YES'' to family planning measures (e.g., condoms and IUDs like copper-T), and the majority of them responded with a ``Yes'' to iron and calcium supplements intake during pregnancy following doctor's advice. These responses highlight a good understanding of the importance of prenatal checkups. When it comes to the health of the children, many mothers rely on Anganwadi workers for immunization and child nutrition education. Their utilization of contraception methods, such as condoms and IUDs like copper-T, demonstrates a proactive approach to managing their reproductive health. 
However, we saw little hesitation and shyness among these women while responding to the questions about family planning as they leaned forward and whispered about their contraceptive methods. Their responses indicate a willingness to engage in conversations about these important issues. Overall, these findings suggest a receptive community towards preventative healthcare and responsible family planning practices. Their trust is placed in these NGOs and healthcare providers in community-based relationships with the community workers who reach out to them from time to time to make them feel heard and bring their feedback to help improve these programs.

We also noticed a difference in the engagement between the participants from the two NGOs. The participants who were recruited through NGO1 and interviewed in person in the comfort of their homes were less engaged as compared to the participants in the focus groups. The focus groups took place at NGO2's center where they bring their disabled kids for various learning activities to improve mobility. These participants were more enthusiastic about the chatbots and new information. In contrast, the participants at home were hesitant to speak and were distracted by their kids and family members. High interest and curiosity among the women in the focus groups may be due to the health condition of their kids, and because they felt safe in that space to discuss their health choices and were not distracted by other responsibilities at home.
Another factor that could have affected their level of engagement could be the frequency and in-person contact with the NGO. Participants in the FGDs had regular, daily contact with NGO2's staff, which likely fostered a stronger rapport and engagement during these discussions. These participants were particularly interested in topics related to their children's diet and developmental milestones. On the other hand, the participants who were connected with only weekly calls and messages had limited face-to-face interaction with NGO1's workers, which may have influenced their engagement during the interviews. This highlights the importance of regular one-on-one interaction for building trust and engagement in health interventions.

Despite widespread smartphone and WhatsApp use among participants, familiarity with chatbots was low. However, there was a positive sentiment towards using chatbots for health information, suggesting a potential to improve health access when the system is designed to cater to audiences who struggle to adopt them.

%% file: future_work.tex
\section{Discussion}

Our findings reveal critical barriers to the adoption of chatbots among women in underserved communities, which structure their digital interactions. 
Below we reflect on the implications of these barriers in chatbot design and implementation efforts.

Participants shared that their phone engagement is limited to short time intervals, often when their children are sleeping. Another gap is in accessible health information, particularly on child nutrition. Mothers expressed a need for advice on feeding practices for their children. A reliance on traditional sources of health advice, such as family elders, sometimes leads to misinformation or missed opportunities for early intervention. For example, a participant mentioned not recognizing early signs of developmental delays in her child due to this reliance. This calls for more culturally relevant, personalized health information that accounts for the specific needs of the child and family context.

Cultural and gender norms additionally amplify these challenges. Family dynamics play a role in decision-making around health and technology use, particularly in communities where women often rely on elders or husbands. These norms not only influence decisions about family planning and healthcare but also limit women's ability to use technology independently, reflecting deeply rooted power dynamics. Technology solutions must take these social structures into account to promote acceptance and use.

Digital literacy also appeared as a limiting factor. Some participants struggled with typing and reading in English and Hindi, which reduced their confidence in navigating these digital health platforms. Many prefer audio or visual content over text-based communication, as it allows them to access information without reading or typing. For chatbots to be effective in such contexts, features like text-to-speech, voice commands, or multimedia content should be prioritized to ensure inclusivity.

The participants in this study significantly trusted the NGO programs they were involved in, particularly when receiving health-related advice. These programs are well-received due to their personalized and accessible weekly messages. This trust was built over time through regular communication and support from local community health workers. Collaborating with such trusted organizations and integrating feedback from participants will be essential for the success during deployment of such technologies.
These barriers and mothers' interest in accessing health information for their children suggest that digital health interventions, such as chatbots, must offer asynchronous, concise, and personalized interactions and accessible solutions that fit into the everyday lives and priorities of women, who have many responsibilities and limited time.

Finally, we wish to highlight that many of these challenges are not unique to chatbots, and have been uncovered in past literature on digital health. But it is worth examining why these foundational issues still persist. Our work speaks to deeper issues around women's agency and their care burden, which has seen limited progress even as digital access has improved, and global health organizations push for digital health solutions. There is a need to address women's control over their own finances to be able to improve digital access, and a society-wide cultural change in seeing their role as more than mere care providers, to have any long-lasting change and improvements in community health outcomes. Otherwise, even as new technologies are introduced, global health and technology practitioners and researchers will continue to experience the same challenges that the field has experienced thus far.